\documentclass[conference, a4paper]{IEEEtran}

\usepackage{cite}
\usepackage{amsmath,amssymb,amsfonts}
\usepackage{graphicx}
\usepackage{textcomp}
\usepackage{xcolor}
\def\BibTeX{{\rm B\kern-.05em{\sc i\kern-.025em b}\kern-.08em
    T\kern-.1667em\lower.7ex\hbox{E}\kern-.125emX}}

\usepackage{mathrsfs}
\usepackage{stfloats}
\usepackage{subfigure}
\usepackage{dsfont}
\usepackage{setspace}
\usepackage{array}
\usepackage{bbm}
\usepackage{hyperref}
\usepackage{bm}
\usepackage{comment}
\usepackage{balance}
\usepackage{acronym,xcolor}

\usepackage{accents}

\usepackage{xcolor}
\usepackage{epstopdf}

\usepackage{algorithm,algorithmic}

\allowdisplaybreaks[4]    

\acrodef{$P_{EM}$}{probability of emulation, or false alarm}
\acrodef{$P_{FA}$}{probability of false alarm}
\acrodef{$P_{MD}$}{probability of missed detection}
\acrodef{$P_{D}$}{probability of detection}

\acrodef{ACF}{autocorrelation function}
\acrodef{ACG}{automatic	gain control}
\acrodef{ACI}{adjacent channel interference}
\acrodef{ACK}{acknowledge}
\acrodef{AcR}{autocorrelation receiver}
\acrodef{ADC}{analog-to-digital converter}
\acrodef{AF}{amplify \& forward}
\acrodef{AFL}{anchor-free localization}
\acrodef{AGNSS}{assisted-GNSS}
\acrodef{AGPS}{assisted GPS}
\acrodef{AI}{automatic identification}
\acrodef{AIC}{Akaike information criterion}
\acrodef{AoA}{angle-of-arrival}
\acrodef{AoD}{angle-of-departure}
\acrodef{AOT}{approximate optimum threshold}
\acrodef{AP}{access point}
\acrodef{API}{application programming interface}
\acrodef{ASK}{amplitude shift keying}
\acrodef{ASNR}{accumulated signal-to-noise ratio}
\acrodef{AUB}{asymptotic union bound}
\acrodef{AWGN}{additive white Gaussian noise}
\acrodef{BAN}{body area network}
\acrodef{BAV}{balanced antipodal Vivaldi}
\acrodef{BCH}{Bose Chaudhuri Hocquenghem}
\acrodef{BEP}{bit error probability}
\acrodef{BER}{bit error rate}
\acrodef{BF}{brute force}
\acrodef{BFC}{block fading channel}
\acrodef{BIC}{Bayesian information criterion}
\acrodef{BLUE}{best linear unbiased estimator}
\acrodef{BPAM}{binary pulse amplitude modulation}
\acrodef{BPF}{bandpass filter}
\acrodef{BPPM}{binary pulse position modulation}
\acrodef{bps}{bits per second}
\acrodef{BPSK}{binary phase shift keying}
\acrodef{BPZF}{band-pass zonal filter}
\acrodef{BS}{base station}
\acrodef{BSC}{binary symmetric channel}
\acrodef{BTB}{Bellini-Tartara bound}
\acrodef{c.c.d.f.}{complementary cumulative distribution function}
\acrodef{c.d.f.}{cumulative distribution function}
\acrodef{CAD}{computer-aided design}
\acrodef{CAIC}{consistent Akaike information criterion}
\acrodef{CAP}{continuous aperture phased}
\acrodef{CCF}{cross correlation function}
\acrodef{CCI}{co-channel interference}
\acrodef{CD}{cooperative diversity}
\acrodef{CDMA}{code division multiple access}
\acrodef{CEOT}{channel ensemble optimum threshold}
\acrodef{CEP}{codeword error probability}
\acrodef{CFAR}{constant	 false alarm rate}
\acrodef{ch.f.}{characteristic function}
\acrodef{CH}{cluster head}
\acrodef{CIR}{channel impulse response}
\acrodef{CL}{centroid localization}
\acrodef{CM}{channel model}
\acrodef{CNR}{clutter-to-noise ratio}
\acrodef{CP}{ciclic prefix}
\acrodef{CPR}{channel pulse response}
\acrodef{CR}{channel response}
\acrodef{CRB}{Cram\'{e}r-Rao bound}
\acrodef{CRC}{cyclic redundancy check}
\acrodef{CRLB}{Cram\'{e}r-Rao lower bound}
\acrodef{CS}{clock skew}
\acrodef{CSCG}{circularly symmetric complex Gaussian}
\acrodef{CSI}{channel state information}
\acrodef{CSMA}{carrier sense multiple access}
\acrodef{CSS}{chirp spread spectrum}
\acrodef{CTS}{clear-to-send}
\acrodef{CW}{continuous wave}
\acrodef{DAA}{detect and avoid}
\acrodef{DAB}{digital audio broadcasting}
\acrodef{DBB}{digital base band}
\acrodef{DBPSK}{differential binary phase shift keying}
\acrodef{DCM}{dual-carrier modulation}
\acrodef{DDP}{detected direct path}
\acrodef{DF}{detect \& forward}
\acrodef{DFMS}{monopole dual feed stripline antenna}
\acrodef{DGPS}{differential GPS}
\acrodef{DLL}{delay-locked loop}
\acrodef{DoD}{Department of Defense}
\acrodef{DoF}{degrees of freedom}
\acrodef{DP}{direct path}
\acrodef{DR}{detection rate}
\acrodef{DRT}{distance ratio test}
\acrodef{DS-SS}{direct-sequence spread-spectrum}
\acrodef{DS}{delay spread}
\acrodef{DTR}{differential transmitted-reference}
\acrodef{DVB-H}{digital video broadcasting\,--\,handheld}
\acrodef{DVB-T}{digital video broadcasting\,--\,terrestrial}
\acrodef{e.m.}{electromagnetic}
\acrodef{ECC}{European Community Commission}
\acrodef{ED}{energy detector}
\acrodef{EDR}{energy detector receiver}
\acrodef{EFIM}{equivalent Fisher information matrix}
\acrodef{EIRP}{effective radiated isotropic power}
\acrodef{EKF}{extended Kalman filter}
\acrodef{ELP}{equivalent low-pass}
\acrodef{EM}{electromagnetic}
\acrodef{EMCB}{extended Miller Chang bound}
\acrodef{EME}{minimum eigenvalue ratio detector}
\acrodef{ENP}{estimated noise power}
\acrodef{ESA}{European Space Agency}
\acrodef{EU}{European Union}
\acrodef{FAR}{false alarm rate}
\acrodef{FCC}{Federal Communications Commission}
\acrodef{FDMA}{frequency division multiple access}
\acrodef{FDMA}{frequency division multiple access}
\acrodef{FEC}{forward error correction}
\acrodef{FEC}{forward error correction}
\acrodef{FFD}{full function device}
\acrodef{FFR}{full function reader}
\acrodef{FF}{far-field}
\acrodef{FFT}{fast Fourier transform}
\acrodef{FG}{factor graph}
\acrodef{FH-SS}{frequency-hopping spread-spectrum}
\acrodef{FH}{frequency-hopping}
\acrodef{FIM}{Fisher information matrix}
\acrodef{FLL}{Frequency-locked loop}
\acrodef{FS}{frame synchronization}
\acrodef{GA}{Gaussian approximation}
\acrodef{GD}{gradient descent}
\acrodef{GDOP}{geometric dilution of precision}
\acrodef{GLR}{generalized likelihood ratio}
\acrodef{GLRT}{generalized likelihood ratio test}
\acrodef{GML}{generalized maximum likelihood}
\acrodef{GPRS}{general packet radio service}
\acrodef{GPS}{global positioning system}
\acrodef{HAP}{high altitude platform}
\acrodef{HCRB}{hybrid Cram\'{e}r-Rao bound}
\acrodef{HDSA}{high-definition situation-aware}
\acrodef{Hi-RADIAL}{High-accuracy RAdio Detection, Identification, And Localization}
\acrodef{HMM}{hidden Markov model}
\acrodef{HPA}{high-power amplifier}
\acrodef{HPBW}{half power beam width}
\acrodef{HW}{hardware}
\acrodef{i.i.d.}{independent, identically distributed}
\acrodef{ICT}{information and communication technologies}
\acrodef{IE}{informative element}
\acrodef{IEEE}{Institute of Electrical and Electronics Engineers}
\acrodef{IF}{intermediate frequency}
\acrodef{IFFT}{inverse fast Fourier transform}
\acrodef{IMF}{ideal matched filter}
\acrodef{IMU}{inertial measurement unit}
\acrodef{INR}{interference-to-noise ratio}
\acrodef{INS}{inertial navigation system}
\acrodef{IoT}{Internet of things}
\acrodef{IIoT}{industrial Internet of things}
\acrodef{INS}{inertial navigation system}
\acrodef{IR-UWB}{impulse radio UWB}
\acrodef{IR}{impulse radio}
\acrodef{IRI}{inter-reader interference}
\acrodef{IRS}{intelligent reflecting surface} 
\acrodef{ISI}{inter-symbol interference} 
\acrodef{isi}{intra-symbol interference} 
\acrodef{ISM}{industrial, scientific and medical}
\acrodef{ISNR}{interference-plus-signal-to-noise-ratio}
\acrodef{IT}{interference temperature}
\acrodef{ITC}{information theoretic criteria}
\acrodef{JBSF}{jump back and search forward}
\acrodef{JF}{just forward}
\acrodef{KF}{Kalman filter}
\acrodef{LDC}{low duty cycle}
\acrodef{LDPC}{low density parity check}
\acrodef{LEO}{localization error outage}
\acrodef{LIS}{large intelligent surface}
\acrodef{LLR}{log-likelihood ratio}
\acrodef{LLRT}{log-likelihood ratio test}
\acrodef{LRT}{likelihood ratio test}
\acrodef{LNA}{low-noise amplifier}
\acrodef{LOS}{line-of-sight}
\acrodef{LRT}{likelihood ratio test}
\acrodef{LS}{least square}
\acrodef{LS}{least squares}
\acrodef{M-PSK}{$M$-ary phase shift keying}
\acrodef{M-QAM}{$M$-ary quadrature amplitude modulation}
\acrodef{m.g.f.}{moment generating function}
\acrodef{MAC}{medium access control}
\acrodef{MAE}{mean absolute error}
\acrodef{MAI}{multiple access interference}
\acrodef{MAN}{metropolitan area network}
\acrodef{MAP}{maximum a posteriori}
\acrodef{MB-OFDM}{multi-band OFDM}
\acrodef{MB-UWB}{multi-band UWB}
\acrodef{MB}{multi-band}
\acrodef{MC}{multi-carrier}
\acrodef{MCB}{Miller Chang bound}
\acrodef{MCRB}{modified Cram\'{e}r-Rao bound}
\acrodef{MDD}{minimum distance distribution}
\acrodef{MDL}{minimum description length}
\acrodef{MF}{matched filter}
\acrodef{MGF}{moment generating function}
\acrodef{MI}{mutual information}
\acrodef{MIMO}{multiple-input multiple-output}
\acrodef{MISO}{multiple-input single-output}
\acrodef{ML}{maximum likelihood}
\acrodef{MM}{min-max}
\acrodef{MME}{maximum-minimum eigenvalue ratio detector}
\acrodef{MMSE}{minimum mean-square error}
\acrodef{MPC}{multipath component}
\acrodef{MRC}{maximal ratio combiner}
\acrodef{MS}{mobile station}
\acrodef{MSB}{most significant bit}
\acrodef{MSE}{mean square error}
\acrodef{MSE}{mean squared error}
\acrodef{MSK}{minimum shift keying}
\acrodef{MUI}{multi-user interference}
\acrodef{MUR}{multistatic radar}
\acrodef{MVU}{minimum variance unbiased}
\acrodef{MZZB}{modified Ziv-Zakai bound}
\acrodef{NB}{narrowband}
\acrodef{NBI}{narrowband interference}
\acrodef{NEO}{navigation error outage}
\acrodef{NFER}{near-Þeld electromagnetic ranging}
\acrodef{NF}{near-field}
\acrodef{NFF}{near-field focused}
\acrodef{NL}{nonlinear}
\acrodef{NLOS}{non-line-of-sight}
\acrodef{NP}{Neyman-Pearson}
\acrodef{NTIA}{National Telecommunications and Information Administration}
\acrodef{NTP}{network time protocol}
\acrodef{OAM}{orbital angular momentum} 
\acrodef{OC}{optimum combining}
\acrodef{OFDM}{orthogonal frequency division multiplexing}
\acrodef{OOK}{on-off keying}
\acrodef{OP}{outage probability}
\acrodef{OT}{optimum threshold}
\acrodef{P-Max}{$P$-Max}  
\acrodef{p.d.f.}{probability density function}
\acrodef{p.m.f.}{probability mass function}
\acrodef{PA}{power amplifier}
\acrodef{PAM}{pulse amplitude modulation}
\acrodef{PAN}{personal area network}
\acrodef{PAR}{peak-to-average ratio}
\acrodef{PD}{probability of detection}
\acrodef{PDP}{power delay profile}
\acrodef{PE}{probability of emulation}
\acrodef{PEB}{position error bound}
\acrodef{PEP}{packet error probability}
\acrodef{PF}{particle filter}
\acrodef{PFA}{probability of false alarm}
\acrodef{PHY}{physical layer}
\acrodef{PL}{path-loss}
\acrodef{PLL}{phase-locked loop}
\acrodef{PMD}{probability of missed detection}
\acrodef{PN}{pseudo-noise}
\acrodef{ppm}{part-per-million}
\acrodef{PPM}{pulse position modulation}
\acrodef{PR}{pseudo-random}
\acrodef{PRake}{partial rake}
\acrodef{PRF}{pulse repetition frequency}
\acrodef{PRP}{pulse repetition period}
\acrodef{PSD}{power spectral density}
\acrodef{PSEP}{pairwise synchronization error probability}
\acrodef{PSK}{phase shift keying}
\acrodef{PSWF}{prolate spheroidal wave function}
\acrodef{PU}{primary user}
\acrodef{QAM}{quadrature amplitude modulation}
\acrodef{QoS}{quality of service}
\acrodef{QPSK}{quadrature phase shift keying}
\acrodef{R.V.}{random variable}
\acrodef{RADAR}{radar}
\acrodef{RCS}{radar cross section}
\acrodef{RDL}{"random data limit"}
\acrodef{REM}{radio environment map}
\acrodef{REO}{ranging error outage}
\acrodef{RF}{radio-frequency}
\acrodef{RFID}{radio-frequency identification}
\acrodef{RFR}{reduced function reader}
\acrodef{RFT}{reduced function tag}
\acrodef{RII}{ranging information intensity}
\acrodef{RIS}{reconfigurable intelligent surface}
\acrodef{rms}{root mean square}
\acrodef{RMSE}{root-mean-square error}
\acrodef{ROC}{receiver operating characteristic}
\acrodef{RRC}{root raised cosine}
\acrodef{RSN}{radar sensor network}
\acrodef{RSS}{received signal strength}
\acrodef{RSSI}{received signal strength indicator}
\acrodef{RTLS}{real time locating systems}
\acrodef{RTT}{round-trip time}
\acrodef{S-V}{Saleh-Valenzuela}
\acrodef{SA}{simulated annealing}
\acrodef{SaG}{stop-and-go}
\acrodef{SBS}{serial backward search}
\acrodef{SBSMC}{serial backward search for multiple clusters}
\acrodef{SCM}{supply chain management}
\acrodef{SCR}{signal-to-clutter ratio}
\acrodef{SEP}{symbol error probability}
\acrodef{SIS}{small intelligent surface}
\acrodef{SFD}{start frame delimiter}
\acrodef{SIMO}{single-input multiple-output}
\acrodef{SINR}{signal-to-interference plus noise ratio}
\acrodef{SIR}{signal-to-interference ratio}
\acrodef{SISO}{single-input single-output}
\acrodef{SNR}{signal-to-noise ratio}
\acrodef{SoC}{system on chip}
\acrodef{SoO}{signal of opportunity}
\acrodef{SoP}{system on package}
\acrodef{SOT}{sub-optimum threshold}
\acrodef{SPAWN}{sum-product algorithm over a wireless network}
\acrodef{SPEB}{squared position error bound}
\acrodef{SPMF}{single-path matched filter}
\acrodef{SQNR}{signal-to-quantization-noise ratio}
\acrodef{SS}{spread spectrum}
\acrodef{ST}{simple thresholding}
\acrodef{SU}{secondary user}
\acrodef{SVD}{singular value decomposition}
\acrodef{SW}{software}
\acrodef{SW}{sync word}
\acrodef{TDE}{time delay estimation}
\acrodef{TDL}{tapped delay line}
\acrodef{TDMA}{time division multiple access}
\acrodef{TDOA}{time difference-of-arrival}
\acrodef{TH}{time-hopping}
\acrodef{TNR}{threshold-to-noise ratio}
\acrodef{TOA}{Time-of-arrival}
\acrodef{TOF}{time-of-flight}
\acrodef{TPC}{transmit power control}
\acrodef{TR}{transmitted-reference}
\acrodef{TS}{tabu search}
\acrodef{UAV}{unmanned aerial vehicle}
\acrodef{UB}{union bound}
\acrodef{UDP}{undetected direct path}
\acrodef{UHF}{ultra-high frequency}
\acrodef{ULA}{uniform linear array}
\acrodef{ULP}{user location protocol}
\acrodef{UMP}{uniformly most powerful}
\acrodef{UMPI}{uniformly most powerful invariant}
\acrodef{UT}{user terminal}
\acrodef{UTC}{coordinated universal time}
\acrodef{UTM}{universal transverse Mercator}
\acrodef{UTRA}{UMTS terrestrial radio access}
\acrodef{UAV}{unmanned aerial vehicle}
\acrodef{UUV}{unmanned underwater vehicle}
\acrodef{UWB}{ultrawide-band}
\acrodef{UWBcap}[UWB]{Ultrawide band}
\acrodef{VFIL}{virtual force iterative localization}
\acrodef{VGA}{variable-gain amplifier}
\acrodef{VNA}{vector network analyzer}
\acrodef{WAF}{wall attenuation factor}
\acrodef{WB}{wideband}
\acrodef{WBI}{wideband interference}
\acrodef{WCL}{weighted centroid localization}
\acrodef{WED}{wall extra delay}
\acrodef{WiMAX} {worldwide interoperability for microwave access}
\acrodef{WLAN}{wireless local area network}
\acrodef{WLS}{weighted least squares}
\acrodef{WMAN}{wireless metropolitan area network}
\acrodef{WPAN}{wireless personal area networks}
\acrodef{WRAPI}{wireless research application programming interface}
\acrodef{WSN}{wireless sensor network}
\acrodef{WSR}{wireless sensor radar}
\acrodef{WSS}{wide-sense stationary}
\acrodef{WWB}{Weiss-Weinstein bound}
\acrodef{WWLB}{Weiss-Weinstein lower bound}
\acrodef{ZZB}{Ziv-Zakai bound}
\acrodef{ZZLB}{Ziv-Zakai lower bound}

\newcommand{\lr}{L_R}
\newcommand{\lt}{L_T}

\begin{document}

\title{\huge LoS MIMO-Arrays vs. LoS MIMO-Surfaces
\author{\IEEEauthorblockN{
Marco Di Renzo\IEEEauthorrefmark{1},
Davide Dardari\IEEEauthorrefmark{2}, and
Nicol\`{o} Decarli\IEEEauthorrefmark{3}
}                                     
\IEEEauthorblockA{\IEEEauthorrefmark{1}%
Universit\'e Paris-Saclay, CNRS, CentraleSup\'elec, Laboratoire des Signaux et Syst\`emes, 91192 Gif-sur-Yvette, France}
\IEEEauthorblockA{\IEEEauthorrefmark{2}%
University of Bologna, Department of Electrical, Electronic, Information Engineering (DEI), 40126 Bologna, Italy}
\IEEEauthorblockA{\IEEEauthorrefmark{3}%
Italian National Research Council, Institute of Electronics, Computer, Telecommun. Engineering, 40136 Bologna, Italy}
\IEEEauthorblockA{\emph{marco.di-renzo@universite-paris-saclay.fr}}
}

}

\maketitle

\begin{abstract}
The wireless research community has expressed major interest in the sub-terahertz band for enabling mobile communications in future wireless networks. The sub-terahertz band offers a large amount of available bandwidth and, therefore, the promise to realize wireless communications at optical speeds. At such high frequency bands, the transceivers need to have larger apertures and need to be deployed more densely than at lower frequency bands. These factors proportionally increase the far-field limit and the spherical curvature of the electromagnetic waves cannot be ignored anymore. This offers the opportunity to realize spatial multiplexing even in line-of-sight channels. In this paper, we overview and compare existing design options to realize spatial multiplexing in line-of-sight multi-antenna channels.
\end{abstract}

\section{Introduction}\label{sec:Intro}
The wireless research community has recently expressed major interest in investigating the opportunities offered by the sub-terahertz (sub-THz) band (30-300 GHz) for future mobile communications \cite{WanGaoDiRAlo:21}. At these high frequencies, point-to-point multiple-input multiple-output (MIMO) links can support the transmission of multiple data streams even in line-of-sight (LoS) channels, by capitalizing on the large aperture of the transceivers, the short transmission range between them, and the small wavelength that characterizes sub-THz signals \cite{DarDec:J21}, \cite{SanDAmDeb:J22}. The transmission of multiple data streams on the same physical resource is usually referred to as spatial multiplexing, multimode communication, or high-rank transmission. At sub-THz frequencies, the electromagnetic waves exhibit a distinct spherical wavefront, which shifts traditional design paradigms based on far-field beamforming antenna-arrays towards near-field focused electrically-large surfaces \cite{HuaEtAl:20}, \cite{MDR__Elsevier_2022}, offering the opportunity of integrating communication and radar sensing in a single transceiver as well \cite{ZhaEtAl:J22}.

LoS MIMO communication is not a new field of research and several works exist in the literature, e.g., \cite{Mil:J00}, \cite{TorMadRod:J11}, \cite{DoLeeLoz:J21}. However, conventional spatial multiplexing MIMO schemes have inherently relied on the underlying existence of rich-scattering channels. In point-to-point MIMO links characterized by large-aperture transceivers (also known as extra-large MIMO or XL-MIMO), short distances among the antenna arrays (short-range MIMO), and high frequencies, however, multipath propagation may not be rich enough but, at the same time, multipath propagation becomes not essential to support multimode communications. In light of this emerging trend in wireless communications, which is fueled by the interest in utilizing the sub-THz frequency band, in this article we describe the main existing options to realize spatial multiplexing LoS MIMO communications, and discuss their advantages and limitations. For brevity, we focus our attention on high signal-to-noise ratio (SNR) scenarios, where spatial multiplexing is the desired choice. In general, it is known that the rank of the channel needs to be optimized as a function of the SNR \cite{ChiRim:C00}.

\begin{figure}[!t]
\centering 
\centering\includegraphics[width=\columnwidth]{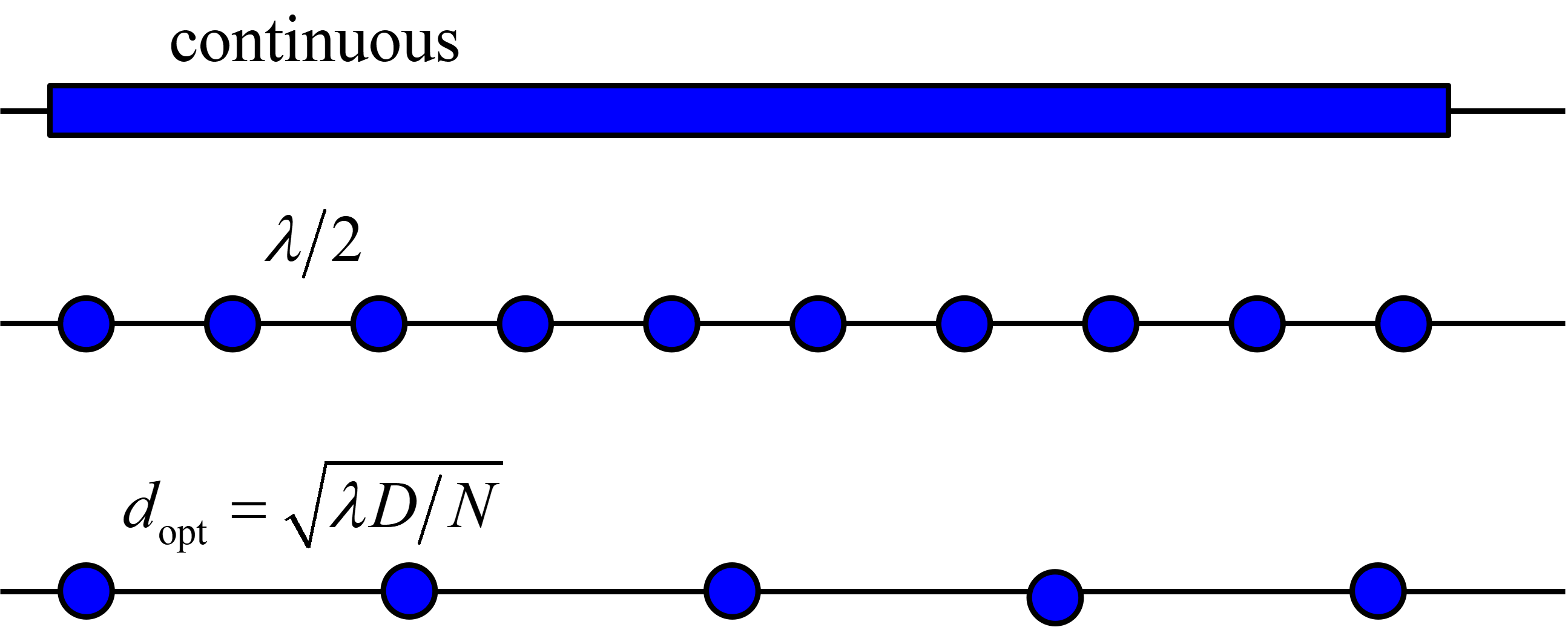}
\caption{Possible options to realize spatial multiplexing in LoS MIMO channels: (top) MIMO-surfaces; (center) half-wavelength spaced MIMO-arrays; (bottom) optimally-spaced MIMO-arrays.} 
\end{figure}
Specifically, we discuss three main design principles and architectures (see Fig. 1): (1) MIMO-surfaces (also known as holographic surfaces) that are virtually continuous electromagnetic objects whose array elements are spaced less than half-wavelength; (2) half-wavelength spaced MIMO-arrays, which is nowadays the typical implementation; and (3) optimally-spaced MIMO-arrays, in which the element spacing is larger than half-wavelength. For ease of discussion, we refer to linear antenna-arrays and, therefore, we consider MIMO-lines instead of MIMO-surfaces. MIMO-lines and MIMO-surfaces are tightly intertwined \cite{Mil:J00}, \cite{ThaMarKarFri:03}. With the term \textit{antenna} we refer to each of the three implementations, considering the antenna-array as a whole electromagnetic object.

\section{Spatial Multiplexing in LoS MIMO Channels}
The possibility of supporting spatial multiplexing in LoS MIMO channels depends on the interplay of
\begin{enumerate}
\item large-aperture antennas
\item short-range links
\item high carrier frequencies
\end{enumerate}
and, specifically, on their relationship according to the concept of Fraunhofer distance of an antenna, which determines the boundary between the far-field region and the radiating near-field region (i.e., the Fresnel region). 

Conventionally, the space around a transmit antenna is divided into several regions, which depend on the characteristics of the field emitted by the antenna itself. The closest area to the antenna corresponds to the reactive near-field region, where the reactive components of the field are dominant. By increasing the observation distance from the antenna, there is the radiating near-field region (or Fresnel region), and then the far-field region (or Fraunhofer region). The boundary of the radiating near-field region and the  far-field region is conventionally identified by the so-called Fraunhofer distance, which is defined as \cite{Tse}
\begin{equation}\label{eq:Fra}
r_{\text{ff}}=\frac{2L^2}{\lambda}
\end{equation}
where $L$ is the antenna size, and $\lambda$ is the wavelength. For antenna-arrays composed of several array elements, $L$ is to be intended as the whole length (aperture) of the antenna.

When two antennas are in the far-field of each other, the rank ($R$) of the corresponding LoS MIMO channel is $R=1$ \cite{Tse}. Therefore, a single communication mode (a plane wave) is well-coupled between the two antennas. In the Fresnel region, on the other hand, the number of communication modes can be larger than one, thus supporting spatial multiplexing even in LoS MIMO channels \cite{Mil:J00}.

The near-field and far-field regions are usually referred to a transmitting antenna \cite{DecDar:J21}. From the point of view of a communication system, however, the operating regions of both the transmitting and receiving antennas need to be considered for correctly determining the possibility of supporting spatial multiplexing in LoS MIMO channels. Consider, for example, the case study of the communication between a small-size antenna and a large-size antenna, and assume that the large-size antenna is in the far-field region of the small-size antenna, while the small-size antenna is within the radiating near-field region of the large-size antenna. This scenario can support the transmission of multiple communication modes (spatial multiplexing), as described in \cite{DecDar:J21}.

In the following section, for ease of presentation, we assume that the paraxial approximation holds true, i.e., the transmitting and receiving MIMO antennas are parallel to each other, their centers are aligned, and they communicate in the Fresnel near-field region \cite{ThaMarKarFri:03}, \cite{Dar:J20}.

\section{LoS MIMO: Models and Architectures}
In this section, we discuss three main design principles and architectures to support spatial multiplexing in LoS MIMO channels (see Fig. 1): (1) MIMO-lines that are virtually continuous electromagnetic objects whose array elements are spaced less than half-wavelength; (2) half-wavelength spaced MIMO-arrays, which is nowadays the \textit{de facto} implementation; and (3) optimally-spaced MIMO-arrays, in which the element spacing is larger than half-wavelength.

\subsection{MIMO-Lines}

As a first option, we consider highly-flexible antennas that are realized, for example, with metasurfaces \cite{MDR__Elsevier_2022}. These structures can approximate ideal apertures, and possess almost complete control and design of the electrical currents on their surface when they operate as transmitters, and complete control in terms of sensing the incident electromagnetic fields when they operate as receivers. In the recent literature, these structures are referred to as \textit{holographic} MIMO \cite{SanDAmDeb:J22}, \textit{continuous aperture} MIMO \cite{SayBeh:C10}, or \textit{large intelligent surfaces} \cite{HuRusEdf:18}.  

As far as MIMO-lines are concerned, the number of communication modes that can be supported in LoS channels is theoretically infinite, since they operate in an infinite-dimensional space. In practice, due to the finite size of the surfaces, only a finite number of modes are, however, strongly coupled. Specifically, the electromagnetic waves that correspond to strongly coupled modes reach the receiving MIMO-line with sufficient power. The electromagnetic waves that correspond to weakly coupled modes, on the other hand, are mostly spread away from the receiving MIMO-line \cite{Mil:J00}. The number of strongly coupled modes is usually referred to as the degrees of freedom (DoF) of the system. In the sequel, this is denoted by $R = \rm{DoF}$.

Under the paraxial propagation condition, the number of strongly coupled communication modes is given, with good approximation, by \cite{Mil:J00}
\begin{equation}\label{eq:Nmodes}
R \approx \max \left\{1,\frac{\lt\lr}{\lambda D}\right\}
\end{equation}
where $\lt$ and $\lr$ are the lengths of the transmitting and receiving MIMO-line, respectively, and $D$ is the distance between their center-points. In particular, \eqref{eq:Nmodes} is valid when $\lt,\lt\ll D$. Differently, when such a condition does not hold, the authors of \cite{DecDar:J21} have shown that the number of strongly coupled communication modes is well approximated by
\begin{align}\label{eq:NParr}
R \approx 1 + \frac{2\lt\lr}{\lambda\sqrt{4D^2+\lr^2}} 
\end{align}
if $\lr\gg\lt$. 

The expression in (3) is applicable in the so-called \textit{geometric near-field region}, which is the operating region in which the distance between the transmitting and receiving antennas is comparable with the their sizes. It is worth mentioning that the geometric near-field region is different from the Fraunhofer far-field distance, since the latter depends on the wavelength as well. Specifically, (3) accounts for the fact that the maximum number of communication modes is upper-bounded by the antenna with the smallest size, i.e., the transmitting surface in (3). Indeed, we obtain $R_{\rm{max}}\to 2\lt/\lambda$ if $\lr \to \infty$.

In order to leverage the $R$ communication modes in (2) or (3), and to transmit $R$ modulated symbols in an interference-free manner, it is necessary to appropriately engineer the electric field (or the surface currents) on the transmitting and receiving antennas. This is sketched in Fig. 2. Specifically, the electric field on the aperture of the two antennas can be expressed in terms of two basis functions ($a_1$, $a_2$, \ldots, $a_{R}$ at the transmitter and $b_1$, $b_2$, \ldots, $b_{R}$ at the receiver, respectively) such that each basis function in the transmitting domain contributes only to one basis function in the receiving domain. The resulting $R$ connected basis functions correspond to the strongly coupled communication modes (whose intensity is denoted by $c_1$, $c_2$, \ldots, $c_{R}$ in Fig. 2). 

In mathematical terms, by virtue of the (scalar) one-dimensional Fresnel-Kirchhoff diffraction integral, we have \cite[Eq. (1)]{ThaMarKarFri:03}
\begin{align}
{E_R}\left( {{x_R}} \right) \propto \int_{{L_T}} {G\left( {{x_T},{x_R}} \right){E_T}\left( {{x_T}} \right)d{x_T}}
\end{align}
with 
\begin{align}
G\left( {{x_T},{x_R}} \right) &= \sum\limits_{n = 1}^\infty  {{c_n}{a_n}\left( {{x_T}} \right){b_n}\left( {{x_R}} \right)} \\
& \approx \sum\limits_{n = 1}^{{R}} {{c_n}{a_n}\left( {{x_T}} \right){b_n}\left( {{x_R}} \right)} 
\end{align}
where ${E_T}\left( {{x_T}} \right)$ and ${E_R}\left( {{x_R}} \right)$ are the electric fields on the aperture of the transmitting and receiving antennas, respectively, and $G\left( {{x_T},{x_R}} \right)$ is the Green function in free space. 

The obtained analytical formulation in (6) is similar to that of spatial multiplexing in conventional MIMO-arrays \cite{Tse}, with the caveat that MIMO-arrays operate in a finite-dimensional space whose dimension is determined by the number of array elements at the transmitting and receiving antennas.
\begin{figure}[!t]
\centering 
\centering\includegraphics[width=\columnwidth]{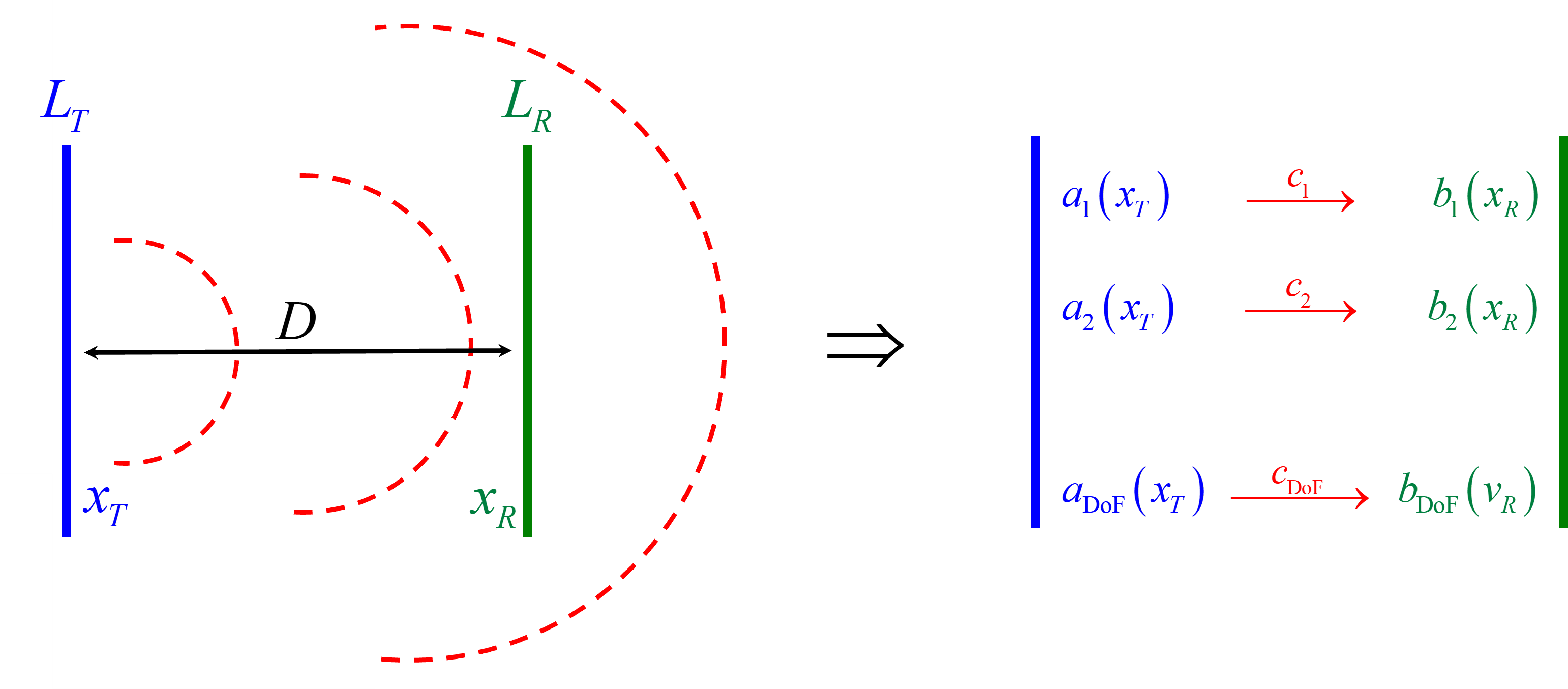}
\caption{A multimode communication channel between two parallel lines can be formulated in terms of basis functions in the transmitting and receiving domains and the corresponding coupling coefficients ($R={\rm{DoF}}$).}
\end{figure}

\subsection{Half-Wavelength Spaced MIMO-Arrays}
In state-of-the-art MIMO communications, the antenna elements of MIMO-arrays are usually spaced at the critical distance of $\lambda/2$. Let us assume that $N$ and $M$ array elements are available at the MIMO transmitter and MIMO receiver, respectively. Let $g_{n,m}$ denote the complex channel between the $n$th element of the transmitting array and the $m$th element of the receiving array, with $n=1,\ldots N$ and $m=1,\ldots M$. The channel elements can be arranged in an $M \times N$ channel matrix $\mathbf{G}=\{g_{n,m}\}$ whose rank ($R$) determines the number of communication modes, and hence the spatial multiplexing gain. By definition, the rank is upper-bounded by
\begin{align}\label{eq:Rankgeneral}
R_{\mathrm{max}} = \min \left\{N,M\right\}.
\end{align}

When the transmitting and receiving MIMO-arrays are in the Fraunhofer far-field region of each other, the channel rank is $R=1$, even if $N,M>1$ \cite{Tse}. If the two MIMO-arrays are in the Fresnel region of each other, on the other hand, the channel rank is $1 \le R \le R_{\mathrm{max}}$. However, the rank cannot be larger than the rank of a MIMO-line, which is approximately equal to \eqref{eq:Nmodes} or \eqref{eq:NParr}. Specifically, from (2) and (3), we have, respectively, the following equations under the paraxial approximation:
\begin{equation}\label{eq:NmodesMIMO}
R \approx \min \left\{ {\max \left\{ {1,\frac{{{L_T}{L_R}}}{{\lambda D}}} \right\},N,M} \right\}
\end{equation}
and
\begin{align}\label{eq:NParrMIMO}
R \approx \min \left\{ {1 + \frac{{2{L_T}{L_R}}}{{\lambda \sqrt {4{D^2} + L_R^2} }},N,M} \right\}.
\end{align}

In the limit $M \rightarrow \infty$ (i.e., $L_R \rightarrow \infty$), the channel rank tends to
\begin{align}
R \approx \min \left\{ {1 + \frac{{2{L_T}}}{\lambda },N} \right\}
\end{align}
which shows that the number of communication modes depends on the electrical length of the MIMO-array with the smallest size.

\subsection{Optimally-Spaced MIMO-Arrays}
A third option to support spatial multiplexing in LoS MIMO channels is to optimize the locations of the array elements at the transmitting and receiving antennas \cite{TorMadRod:J11}. Let us assume that the transmitting and receiving MIMO-arrays have lengths $L_T$ and $L_R$, respectively, and that they are made of $N=M$ array elements. Assuming that the array elements are evenly distributed, the inter-distances between them at the transmitting and receiving antennas are $d_T = L_T/(N-1)$ and $d_R = L_R/(N-1)$, respectively. 

The rationale behind the design of optimally-spaced MIMO-arrays is to identify the locations of the array elements to ensure that the rank of the channel is $R=N$. A rank-$N$ channel can be established provided that the array elements are equally spaced by the Rayleigh distance \cite{TorMadRod:J11}. Without posing limits on the sizes $\lt$ and $\lr$, the Rayleigh spacing results in the following condition:
\begin{equation}\label{eq:subsamplingN}
{d_T}{d_R} = {{\lambda D}}/{N}\mathop  \Rightarrow \limits^{{d_T} = {d_R} = d_{\rm{opt}}} d_{\rm{opt}} = \sqrt {{{\lambda D}}/{N}}.
\end{equation}

Equation (11) is derived under the assumption that the lengths $L_T$ and $L_R$ can be optimized based on the optimal inter-distance $d_{\rm{opt}}$. If the lengths $L_T$ and $L_R$  are, on the other hand, fixed, the number of communication modes that can be supported while preserving the orthogonality is \cite{TorMadRod:J11}
\begin{align}
{R_{\max }} &= 1 + \frac{{{L_T}{L_R}}}{{2\lambda D}} + \sqrt {{{\left( {1 + \frac{{{L_T}{L_R}}}{{2\lambda D}}} \right)}^2} - 1}
\end{align}
\vspace{0.001cm}
\begin{align}
&  \hspace{-0.25cm} \mathop  \approx \limits^{N - 1 \approx N} {\max \left\{ {1,\frac{{{L_T}{L_R}}}{{\lambda D}}} \right\}}. \nonumber
\end{align}

It is interesting to note that (12) is approximately equal to (2), even though the analytical derivation is different. By direct inspection of (12), we note that $d_{\rm{opt}}$ is, in general, greater than $\lambda$, which implies that spatial mutiplexing in LoS conditions requires the deployment of sparse MIMO-arrays.

\subsection{Comparison Among the Three Considered Implementations}
In the previous sub-sections, we have presented three different architectures to support spatial multiplexing in LoS MIMO channels. We have observed that all of them provide the same number of (effective or strongly coupled) communication modes. However, the three architectures have different implementation requirements, and, therefore, complexity and deployment tradeoffs. Specifically, the following considerations are in order.

\textbf{Optimally-spaced MIMO-arrays}. This architecture is able to support a number of communication modes that coincides with the number of antenna elements, provided that no constraints are imposed on the physical size of the array structure. Compared with critically-spaced MIMO-arrays with the same size, a major benefit of optimally-spaced MIMO-arrays is requiring a number of radio frequency chains that is equal to the number of communication modes, which is a parsimonious implementation design \cite{Mil:J00}. However, optimally-spaced MIMO-arrays usually offer a lower beamforming gain compared with critically-spaced MIMO-arrays of the same size, since the latter have a larger number of array elements. This may be an unfavorable feature for transmission at high frequency bands \cite{TorMadRod:J11}. In addition, the optimal inter-distance $d_{\rm{opt}}$ depends on the transmission distance $D$. As a result, the locations of the array elements need to be varied with the transmission distance \cite{ChiRim:C00}, which may not always be possible. Also, the size of the resulting array structure may be large. To circumvent these issues, rotations or the electronic selection of multiple antenna arrays with a radial disposition may be utilized \cite{DoLeeLoz:J21}. 

\textbf{Half-wavelength spaced MIMO-arrays}. Critically spaced MIMO-arrays provide a solution to circumvent the need of varying the locations of the array elements as a function of the transmission distance. The larger number of array elements, in addition, provides larger beamforming gains if the size of the two structures is the same \cite{TorMadRod:J11}. This architectural solution, however, necessitates a number of radio frequency chains that is equal to the number of array elements, which is usually greater than the number of communication modes supported by the LoS MIMO channel. Specifically, we have
\begin{equation}
\frac{{{L_T}{L_R}}}{{\lambda D}} \le \left[ {\left( {\frac{{2{L_T}}}{\lambda }} \right) + \left( {\frac{{2{L_R}}}{\lambda }} \right)} \right]
\end{equation}
where the left-hand side denotes the number of radio frequency chains required for optimally-spaced MIMO arrays and MIMO-lines, and the right-hand side denotes the number of radio frequency chains required for half-wavelength spaced MIMO-arrays.

An option to reduce the number of radio frequency chains consists of utilizing arrays of sub-arrays, in which each sub-array is centered at the location identified by the optimal inter-distance in (11) and the elements of each sub-array are critically spaced at $\lambda/2$. The number of radio frequency chains is equal to the rank of the channel, but the centers of the sub-arrays need to be varied with the transmission distance \cite{TorMadRod:J11}.

\textbf{MIMO-lines}. MIMO-lines enable the design of hybrid MIMO architectures with a number of radio frequency chains that is equal to the rank of the channel, provided that optimal basis functions are utilized for engineering the electric fields on the transmitting and receiving apertures (as illustrated in Fig. 2) \cite{DarDec:J21}. If the number of radio frequency chains is given, in addition, the number of effective communication modes can be inherently adapted to the transmission distance, the apertures of the antennas, and the operating frequency, provided that the basis functions on the transmitting and receiving apertures are appropriately chosen \cite{ThaMarKarFri:03}. Compared to $(\lambda/2)$-spaced MIMO-arrays, the benefits of continuous-aperture MIMO-antennas are well known, and include the possibility of reducing the gratings lobes and providing higher performance for large values of oblique angles of incidence \cite{SayBeh:C10}. Moreover, quasi-continuous sub-wavelength implementations offer the possibility to exploit the mutual coupling among the antenna elements by design, and to realize super-directive MIMO-antennas \cite{Mar:C19}. In fact, $(\lambda/2)$-spaced MIMO-arrays are sufficient to sense an electric field, provided that the observation plane is at distances from the transmitting and receiving antennas where the evanescent waves can be ignored \cite[Section 1.4.3]{yu2015advanced}. Otherwise, it is necessary to consider inter-distances shorter than half-wavelength. However, the critical spacing at $\lambda/2$ renders the mutual coupling among the antenna elements negligible, thus simplifying the antenna design.

\section{Conclusion} 
In this paper, we have overviewed and compared existing design options to support spatial multiplexing in LoS MIMO channels, by considering implementations based on MIMO-arrays and MIMO-surfaces.

\section*{Acknowledgment}
The research work of Marco Di Renzo was in part supported by the European Commission through the H2020 ARIADNE (grant 871464) and RISE-6G (grant 101017011) projects, and the Fulbright Foundation. The research work of Davide Dardari was in part supported by the European Commission through the HE TIMES (grant 101096307) project.

\balance

\bibliographystyle{IEEEtran}
\bibliography{IEEEabrv,StringDefinitions,StringDefinitions2,biblio}

\end{document}